\documentclass[aps,pra,twocolumn,showpacs,superscriptaddress,letterpaper,longbibliography]{revtex4-1}

\usepackage{graphicx}
\usepackage{mdframed}
\usepackage{framed}
\usepackage{indentfirst}
\usepackage{natbib}
\usepackage{amsmath,amssymb}
\usepackage{subfigure}
\usepackage{enumitem}
\usepackage{epstopdf}
\usepackage{xfrac}
\usepackage{multirow}

\begin{document}

\title{Research-based assessment of students' beliefs about experimental physics:\\When is gender a factor?}

\pacs{01.40.Fk}
\keywords{physics education research, upper-division, laboratory, attitudes, assessment, gender}

\author{Bethany R. Wilcox}
\affiliation{Department of Physics, University of Colorado, 390 UCB, Boulder, CO 80309}

\author{H. J. Lewandowski}
\affiliation{Department of Physics, University of Colorado, 390 UCB, Boulder, CO 80309}
\affiliation{JILA, National Institute of Standards and Technology and University of Colorado, Boulder, CO 80309}

\begin{abstract}
The existence of gender differences in student performance on conceptual assessments and their responses to attitudinal assessments has been repeatedly demonstrated.  This difference is often present in students' preinstruction responses and persists in their postinstruction responses.  However, one area in which the presence of gender differences has not been extensively explored is undergraduate laboratory courses.  For example, one of the few laboratory focused research-based assessments, the Colorado Learning Attitudes about Science Survey for Experimental Physics (E-CLASS), has not been tested for the existence of gender differences in students' responses.  Here, we utilize a national data set of responses to the E-CLASS to determine if they demonstrate significant gender differences.  We also investigate how these differences vary along multiple student and course demographic slices, including course level (first-year vs.\ beyond-first-year) and major (physics vs.\ non-physics).  We observe a gender gap in pre- and postinstruction E-CLASS scores in the aggregate data both for the overall score and for most items individually.  However, for some subpopulations (e.g., beyond-first-year students) the size or even existence of the gender gap depends on another dimension (e.g., student major).  We also find that for all groups the gap in postinstruction scores vanishes or is greatly reduced when controlling for preinstruction scores, course level, and student major. 
\end{abstract}

\maketitle

\section{\label{sec:intro}Introduction \& Motivation}

Student learning in laboratory physics courses has emerged as a new and growing area of research within the physics education research (PER) community (e.g., \cite{zwickl2013adlab, volkwyn2008measurement, etkina2001isle}).  Laboratory courses have also been specifically called out as critical pieces of the undergraduate curriculum by professional groups within several disciplines \cite{AAPT2015guidelines, olson2012excel, nrc2003bio}.  Lab courses have garnered this increased attention in part because they represent unique learning environments \cite{zwickl2013adlab}.  These courses are one of the few places, outside of undergraduate research experiences, that can provide students with opportunities to develop the practical lab skills that will help prepare them for a future in industry, teaching, or graduate school.  Lab courses also offer valuable opportunities for students to engage in a range of authentic scientific practices, such as designing and building experiments, collecting and interpreting data, and communicating scientific content.  As such, laboratory course environments represent a key component of helping students to develop expert-like epistemologies and habits of mind, as well as enthusiasm and confidence in research.  


As part of recent laboratory course transformation efforts at the University of Colorado Boulder (CU) \cite{zwickl2013adlab}, Zwickl \emph{et al.} developed a laboratory-focused assessment specifically targeted at the broader, non-content learning goals discussed above.  The assessment, known as the E-CLASS (the Colorado Learning Attitudes about Science Survey for Experimental Physics) \cite{zwickl2014eclass}, is a 30-item, Likert-style survey that includes multiple items targeting students' epistemologies and expectations as to the nature of experimental physics along with several items targeting student affect and confidence when performing physics experiments.  This assessment was intended to be used in both introductory and advanced lab courses and, thus, includes items targeting a wide range of learning goals \cite{zwickl2014eclass}.  Items on the E-CLASS feature a paired structure in which students are prompted with a statement (e.g., ``The primary purpose of doing physics experiments is to confirm previously known results.'')\ and asked to rate their level of agreement on a 5-point Likert scale both from their perspective when doing experiments for class and that of a hypothetical experimental physicist.  E-CLASS was validated through student interviews and faculty review, and has been tested for statistical validity and reliability using responses from a broad student population \cite{wilcox2016eclass}.  See Ref.\ \cite{ECLASSwebsite} for more information about the E-CLASS as well as a list of all question prompts.  

While E-CLASS is the first laboratory-specific assessment of this type, a number of related assessments have been developed for examining students attitudes, beliefs, and epistemologies about physics more generally.  For example, the Maryland Physics Expectation Survey (MPEX) \cite{redish1998mpex} and the Colorado Learning Attitudes about Science Survey (CLASS) \cite{adams2006class} were developed to probe students beliefs and expectations about physics and physics learning before and after completing lecture physics courses.  One notable result from the CLASS was the appearance of significant gender differences in students' responses with women generally providing less expert-like responses \cite{adams2006class, gray2008class}.  Prior work has also demonstrated that students beliefs, as measured by the CLASS, are correllated with both their self-reported interest in physics \cite{perkins2005class, perkins2006class} and their performance on certain conceptual assessments \cite{perkins2004class}.   As both interest and performance are important aspects of a students' persistance in a given major, understanding gender differences in assessments like the CLASS or E-CLASS can be particularly relevant for retention of women within the physics major.  

The observation of gender differences in non-content assessments like the CLASS is complementary to a large body of literature documenting the appearances of a gender gap in scores on content-focused assessments in lecture courses (see Ref.\ \cite{madsen2013gender} for a review).  The origin of this gender gap in these assessments is not well understood and is likely driven by multiple, complex factors.  However, as indicated previously, the appearance and persistence of the gender gap is of particular interest in relation to the underrepresentation of women in physics \cite{ivie2005women}.  Consistently lower performance relative to the men in their classes may be a contributing factor in discouraging women from persisting in the physics major.  However, the existence of the gender gap in laboratory courses assessments has been less well explored.  One notable exception is work by Day \emph{et al.} \cite{day2016gender}, in which they examined students scores on the laboratory assessment known as the Concise Data Processing Assessment (CDPA) \cite{day2011cdpa} with respect to gender differences.  Day \emph{et al.} found a significant gender gap in both the pre- and postinstruction scores on the CDPA; however, they also note that classroom observations of these students provided no indication that female students are less capable of learning than their peers.  

The goal of this paper is to present the first large-scale analysis of students' responses to the E-CLASS with respect to gender differences.  Documentation of known gender differences will provide an important resource for instructors and researchers interested in using the E-CLASS and interpreting the results appropriately.  Here, we review the existing PER literature on gender differences including critiques and defenses of this body of research (Sec.\ \ref{sec:background}).  We also describe the data sources and analysis used for this study (Sec.\ \ref{sec:methods}).  We then present results with respect to gender differences in students' raw pre- and postinstruction E-CLASS scores and gains (Sec.\ \ref{sec:rawResults}) and explore how these differences vary along different demographic lines (e.g., majors vs.\ non-majors) (Sec.\ \ref{sec:stdemResults}).  In addition to examining raw scores and learning gains, we also investigate whether the gender gap in postinstruction scores persists after controlling for preinstruction scores and other factors (Sec.\ \ref{sec:ancovaResults}).  Finally, we end with a discussion of limitations of the study and future work (Sec.\ \ref{sec:discussion}).

\section{\label{sec:background}Background}

\subsection{\label{sec:epistBackground}Epistemology, affect, and labs}

In this section, we discuss the background on, and intersection of, epistemology, affect, gender, and lab courses.  The affective items on the E-CLASS are those that target students' interests, attitudes, emotional responses, and confidence when doing physics experiments \cite{krathwohl1964affect}.  The epistemological items on the E-CLASS are those that target students' theories of the nature of knowledge, knowing, and learning with respect to a particular discipline \cite{zwickl2014eclass, hofer1997epistemology, elby2009epistemology}. For experimental physics, this includes students' views as to what makes a good or valid experiment, and what are the appropriate ways to understand the design and operation of an experiment and the communication of results \cite{zwickl2014eclass}.  We ground our interpretation of students' responses to the E-CLASS in a resources perspective on the nature of epistemological beliefs in which students are expected to draw on a range of resources and experiences when responding to each E-CLASS statement \cite{hammer2005resources}.  Thus, a  student's responses might sometimes be in apparent contradiction with each other due to contextual differences.  This view is as opposed to assuming that students hold coherent and stable epistemological stances based on a well-developed world view of doing physics experiments \cite{zwickl2014eclass}. 

While the relationship between students' gender and their attitudes and beliefs about experimental physics has not been explored previously, multiple studies have demonstrated gender differences in students' attitudes towards, and beliefs about, physics or science more generally.  In addition to the studies described earlier documenting gender differences in CLASS scores within undergraduate physics courses \cite{adams2006class, gray2008class}, there have been similar studies examining attitudinal differences within both lecture and lab courses in other disciplines (e.g., chemistry) and at other educational levels (e.g., high school courses).  For example, Weinburgh \cite{weinburgh1995gender} reviewed 18 studies examining gender differences in students attitudes towards science and found that 81\% of the gendered comparisons included in these studies reported men showing more positive attitudes towards science than women.  Prior work has also repeatedly shown a link between students attitudes and beliefs and both their achievement in science \cite{weinburgh1995gender, freedman2002gender, perkins2004class} and their decision to pursue and/or persist in their scientific education \cite{simpson1990gender}.  Thus, the appearance and consistency of gender differences in students attitudes and beliefs about science generally, and physics specifically, is of particular concern with respect to the underrepresentation of women in these disciplines. As undergraduate lab courses often have an explicit or implicit goal of promoting expert-like epistemologies and habits of mind, as well as enthusiasm and confidence in research \cite{zwickl2013adlab}, it is important to determine if the same gendered trends seen in lecture courses and other disciplines are also observed in the context of lab courses.

\subsection{\label{sec:genderBackground}Gender gap research}

In this section, we review some of the literature  in PER around gender differences or the gender gap, discuss critiques and defenses of this body of literature, and articulate our stance with respect to these issues.

Gender differences in students' performance in physics courses at the undergraduate level have been, and continue to be, a significant focus of the PER community, as evidenced by the recent call from this journal for a focused collection around issues of gender in physics.  Danielsson \cite{danielsson2010gender} reviewed 57 articles related to gender and physics education.  In addition to summarizing the findings of these articles, Danielsson classified a majority of these studies as including characterizations of female students' performance relative to that of male students.  In addition to the studies around gender differences in student responses to the CLASS discussed previously \cite{adams2006class, gray2008class}, there have also been a number of quantitative studies of gender differences in scores on conceptual assessments.  Madsen \emph{et al.} \cite{madsen2013gender} recently reviewed this body of literature through a meta-analysis of 26 published studies documenting the gender gap on research-based assessments.  They found that, while these studies consistently showed a gender gap in students' scores and gains, the size of the gap, how it developed over time, and what factors influenced it varied significantly between studies.  They used these results to conclude that the gender gap was likely due to a combination of multiple factors over time, rather than the result of a single consistent issue.  

While gender gap research in PER has garnered significant attention, there have been a number of critiques of both gender gap literature specifically and performance gap literature more broadly \cite{gutierrez2008gap, martin2009race, traxler2016gender}.  First, one of the major critiques is that gender gap literature treats gender as a strict binary without acknowledging that there are many who do not fit into the distinct and simplified categories of `men' and `women' \cite{traxler2016gender}.  To begin addressing this issues, Traxler \emph{et al.} \cite{traxler2016gender} advocate for a new framework focusing on ``gender performativity.''  In this framework, gender is treated as something that is enacted rather than as a predetermined state \cite{butler1988performative}.  A second critique of performance gap literature is that it implicitly suggests that between-group variance is greater than within-group variance \cite{gutierrez2008gap}.  In other words, it implies that the differences between men and women are larger or more important than the differences between individual women or subgroups of women.    A third critique of the performance gap literature is that it inherently sets up the majority (in this case, men) as defining ``excellence'' \cite{gutierrez2008gap, martin2009race, traxler2016gender}, and generally leads to a deficit model of the underrepresented group.  This deficit model implies that the solution to the performance gap is to ``fix'' the underrepresented group in order to make them more like the majority.  This perspective fails to acknowledge cultural power structures within the education system that work to support the majority group while simultaneously suppressing groups not seen as part of this majority \cite{gutierrez2008gap, martin2009race}.  Finally, performance gap literature has also been critiqued for focusing primarily on the appearance of the gap without addressing or identifying its root cause(s) \cite{gutierrez2008gap}.  

Despite the critiques of performance gap research, there are others who argue that analyses of achievement gaps still represent an important and valuable research area \cite{lubienski2008gap}.  These arguments center on the potential impact of the findings of performance gap studies both in terms of motivating change at a political and administrative level, and helping researchers identify which groups and learning environments can benefit from additional equity-focused research efforts.  Moreover, investigations of the nature and dynamics of the gap can be used to refute claims that the gap is a result of biological differences \cite{lubienski2008gap}.  Recent advances in statistical methods (e.g., hierarchical linear modeling \cite{raudenbush2002hlm} and analysis of covariance \cite{wildt1978ancova}) also allow for more sophisticated and nuanced analyses of gender gap data.  

Considering this literature on the potential impacts and limitations of performance gap literature, we take the stance that there are still many opportunities for investigations of the gender gap to provide useful and valuable information for the PER community, particularly in contexts like lab courses where these gaps have not yet been well studied.  However, we also argue that this literature highlights a number of important issues for researchers to attend to, and explicitly acknowledge, when investigating achievement gaps.  For example, we support conceptualizing gender as a complex and non-binary construct.  However, the logistical constraints of large-scale data collection make it difficult to collect nuanced information about gender in an online survey like the E-CLASS.  Thus, as discussed in Sec.\ \ref{sec:methods}, the analysis here focuses on gender as the binary distinction between men and women.  Additionally, while we do make comparisons between men and women, we also investigate the impact of factors that contribute to within-group variance for both men and women (e.g., student major).  In cases where comparisons between men and women do show a gap, we do not interpret these gaps as representing evidence that women are less capable than men.  Rather, these results should be seen as a guide to identify areas worthy of additional quantitative and qualitative work in order to determine the causes of the gap.

\section{\label{sec:methods}Methods}

In this section, we present the data sources, student and institution demographics, and analysis methods used for this study.  

\subsection{\label{sec:data}Data sources}

Data for this study were drawn from an existing data set consisting of seven semesters of students' responses to the E-CLASS collected between 01/2013 and 5/2016 from multiple institutions across the United States.  These data were collected through the E-CLASS centralized administration system \cite{wilcox2016admin} as part of ongoing research regarding students' epistemologies in the context of course transformation efforts in undergraduate laboratory courses (e.g., Ref.\ \cite{zwickl2013adlab}).  The assessment was administered online both pre- and postinstruction, typically in the first and last week of the course or laboratory section.  In addition to collecting student responses for all courses in the data set, the E-CLASS system also collects basic information about course type, institution, and pedagogy for each course.  

The final, seven-semester data set includes matched pre- and postinstruction data from 130 distinct courses across 75 institutions.  These institutions span a range of different types from 2-year colleges to Ph.D. granting universities (see Table \ref{tab:instTypes}).  Several of these institutions administered E-CLASS in multiple semesters of the same course during data collection.  Thus, the full data set includes matched responses from 206 separate instances of the E-CLASS.  These courses include both first-year (FY) courses and beyond-first-year (BFY) courses (see Table \ref{tab:courseTypes}).

\begin{table}
\caption{Number of institutions of different types for which we have matched pre- \emph{and} postinstruction responses to the E-CLASS.}\label{tab:instTypes}
\begin{ruledtabular}
   \begin{tabular}{ l r r r r r }
       & 2-year & 4-year & Master's & Ph.D. \hspace{5mm}& Total\\
       &\hspace{1mm} college &\hspace{1mm} college &\hspace{1mm} granting &\hspace{1mm} granting & \\
     \hline
     Number of & & & & &\\
     Institutions\hspace{1mm}& 3 & 36 & 8 & 28 & 75 \\
   \end{tabular}
\end{ruledtabular}
\end{table}

\begin{table}[b]
\caption{Number of first-year and beyond-first-year courses in the matched data set.  The number of students in the beyond-first-year courses is smaller in part because of the smaller class sizes typical of more advanced physics labs.  The number of separate instances of the E-CLASS accounts for courses that administered E-CLASS more than once in the 7 semesters of data collection.   }\label{tab:courseTypes}
\begin{ruledtabular}
   \begin{tabular}{ l r r r }
       & Distinct & Separate & \hspace{1mm} Number of \\
       &\hspace{1mm} courses &\hspace{1mm} instances & Students  \\
     \hline
     First-year & 63 & 102 & 5609  \\
     Beyond-first-year \hspace{1mm}& 68 & 104 & 1558  \\
   \end{tabular}
\end{ruledtabular}
\end{table}

Student responses were matched pre- to postinstruction first by student ID number, then by first and last name when student ID matching failed.  In addition to eliminating responses that could not be matched from pre- to post-test, certain responses were identified as invalid and eliminated.  For example, students who did not respond correctly to a filtering question, which prompts students to select ``agree'' (not ``Strongly agree''), were dropped from the data set.  For more information on what constitutes a valid response see Ref.\ \cite{wilcox2016eclass}.  The final matched data set included $N=7167$ students representing a response rate of roughly 40\%.  This response rate based on the estimates of the total enrollment provided by instructors on the course information survey and is only an approximation of the true response rate as enrollment may have fluctuated after the instructor completed the information survey.  The response rates for the pre- and post-tests individually were higher -- between $65-75\%$ \cite{wilcox2016eclass}.  While we have no clear measure of how representative our sample is of the overall population, previous research suggests that lower response rates likely results in an underrepresentation of lower performing students \cite{wilcox2016eclass}.  

Gender data were collected as one of the final questions on the postinstruction E-CLASS.  This question was intentionally placed at the end of the instrument and appeared on a separate page in the online interface in order to avoid the potential for triggering stereotype threat \cite{miyake2010stereotype}.  Historically, the item asking for students' gender was phrased, ``What is your gender?,'' and the possible response options were: female, male, or prefer not to say.  This phrasing conflates the distinct constructs of gender and biological sex, and also treats gender as a strict binary.  Both of these practices have been critiqued in the literature around gender studies (see Sec. \ref{sec:background}), and for the final two semesters of data collection (fall 2015 and spring 2016) the response options were changed to: woman, man, or other (text box provided).   Despite the change in phrasing, we have included these data in the data set as we posit that the vast majority of student respondents would have responded consistently to both versions of the question.  

Roughly 2\% ($N=154$) of the overall population selected either ``Prefer not to say'' or ``Other'' (depending on the semester) in response to the gender item.  An examination of the text entered into the text box associated with the ``Other'' category indicates that some students selected this category inappropriately and entered responses like ``cyborg'' or ``male engineer.''  Thus, given the difficulty inherent in characterizing who is actually represented in the group with unknown or other genders, we have excluded these individuals from our analysis.  For the remainder of the paper, our treatment of gender will be restricted to the binary distinction between men and women; however, we caution that this treatment both conflates the ideas of gender and biological sex, and does not reflect a nuanced and non-binary understanding of gender.  

The gender breakdown of the final, matched data set was 38\% ($N=2751$) women and 59\% ($N=4262$) men.  Racial demographic data are not reported here because these data were collected only in the final two semesters of data collection.  Examination of E-CLASS scores with regards to racial dynamics will be the subject of future work after aggregation of sufficient data.  In addition to gender data, the postinstruction E-CLASS also asked students for their primary major.  Table \ref{tab:major} reports the breakdown of students by major in the matched data set as well as by course level and gender.  The students were provided 15 options for primary major, and we have collapsed these options into four categories -- physics (includes engineering physics), engineering (excludes engineering physics), other science (includes math, biology, chemistry, etc.), and nonscience (includes non-science and open-option).  

It is likely the case that students in the various engineering, other science, and nonscience majors have meaningfully different prior laboratory experiences.  Moreover, these students may take other laboratory courses related to their primary major during their undergraduate career, and while the E-CLASS is specifically phrased to target students epistemologies about \emph{experimental physics}, previous research has not explored whether participation in lab courses from other disciplines significantly impacts students' E-CLASS responses.  This suggests that variations in the prior and ongoing experiences of students in non-physics majors are likely significant; however, we are not able to clearly characterize the nature of these differences given the data currently available along with the large number of courses and institutions in the data set.  Given this, and the physics focus of the E-CLASS, we have chosen to focus our analysis of student major on the binary difference between physics and non-physics majors.    Thus in the following analysis, we further collapse the engineering, other science, and nonscience categories to a single ``non-physics'' group.  

\begin{table}
\caption{Breakdown of students by major in the matched data sets ($N=7167$).  Note: `Physics' includes both physics and engineering physics majors; the `Other Science' category includes (but is not limited to) biology, chemistry, and math majors; and `Non-science' includes both declared non-science majors and students who are open option/undeclared.  }\label{tab:major}
\begin{ruledtabular}
   \begin{tabular}{ l c r r r r}
       & N &Physics&  Engineering& Other & Non- \\
       Group& & & (Non-physics)& Science &  science \\
      \hline
      All & 7167 & 21\% & 25\% & 46\% & 7\% \\
      \hline
      FY & 5609 & 7\% & 27\% & 56\% & 9\% \\
      BFY & 1558 & 71\% & 19\% & 7\% & 1\% \\
      \hline
      Men & 4262 & 28\% & 31\% & 34\% & 6\% \\
      Women & 2751 & 10\% & 17\% & 65\% & 8\% \\
   \end{tabular}
\end{ruledtabular}
\end{table}

\subsection{\label{sec:Analysis}Analysis}

Response options for items on the E-CLASS are given on a 5-point Likert scale (strongly agree to strongly disagree).  For scoring purposes, the responses ``strongly (dis)agree'' and ``(dis)agree'' are collapsed, and students' responses are coded as simply agree, disagree, or neutral.   Students are then given a numerical score based on whether their selection is consistent with the established expert-like response: $+1$ point for favorable, $0$ points for neutral, and $-1$ point for unfavorable.  A student's overall score on the assessment is given by the sum of their scores on each of the 30 E-CLASS items resulting in a possible range of scores of $[-30,30]$.  For more information on the scoring of the E-CLASS see Ref. \cite{wilcox2016eclass}.  

Throughout this paper, we will discuss pre- and postinstruction scores as well as learning gains on the E-CLASS both overall and by-item.  In previous work, we have cautioned instructors using the E-CLASS against focusing exclusively on the overall score when interpreting their results \cite{wilcox2016eclass}.  The E-CLASS targets a range of learning goals some of which may not be relevant to a specific course, and we encourage instructors to focus also on the individual items most relevant to their learning goals.  For this reason, we provide a breakdown of gender differences in students' scores by item.  However, the overall score is still useful in that it provides a continuous variable that offers a wholistic view of students' performance on the E-CLASS that can be used to quantitatively examine how that performance varies across subpopulations of students.  As the distribution of E-CLASS scores is typically non-normal (see Sec. \ref{sec:rawResults}), we utilized the non-parametric Mann-Whitney U-test \cite{mann1947mwu} to establish the statistical significance of differences between means of different distributions.  For statistically significant differences, we also report Cohen's d \cite{cohen1988d} as a measure of effect size and practical significance.  The importance of reporting effect size along with statistical significance has been highlighted previously in the context of equity related studies \cite{rodriguez2012equity}.  

Consistent with recommendations by Day \emph{et al.} \cite{day2016gender}, we calculate multiple learning gains (e.g., normalized change, Hake gain, etc.) in order to compare across different metrics (Sec.\ \ref{sec:rawResults}).  Informed by analysis of raw scores and learning gains, we also utilize an analysis of covariance (ANCOVA) \cite{wildt1978ancova} as a method for testing the difference between postinstruction means while accounting for the variance associated with other factors, in this case, preinstruction scores, student major, and course level.  These variables were selected based on prior analysis \cite{wilcox2016eclass, wilcox2015eclass} and our own experience, which suggested they could account for significant amounts of the variance in postinstruction E-CLASS scores.  In order for the results of an ANCOVA to be valid, the data must meet several assumptions.  The assumptions of an ANCOVA are discussed in detail in Refs. \cite{wildt1978ancova, day2016gender}; tests of the E-CLASS matched data showed that they satisfied these assumptions with two exceptions.  In our data, the covariate (i.e., preinstruction score) is not independent of the other variables (i.e., gender, major, and course level). Shared variance between the covariate and independent variables is to be expected in any observational study in which randomized assignment to experimental groups was not done or not possible \cite{miller2001ancova}. Violation of the assumption of covariate independence implies that our results should be interpreted as a lower bound on the relationship between each gender and postinstruction E-CLASS score.  The second violation of the assumptions of ANCOVA is discussed in Sec.\ \ref{sec:ancovaResults}.

\section{\label{sec:Results}Results}

This section presents findings with respect to gender differences on the E-CLASS using raw scores, learning gains, and ANCOVA.  

\subsection{\label{sec:rawResults}Gender differences in the aggregate data}

\begin{figure}[b]
\includegraphics[width=\linewidth]{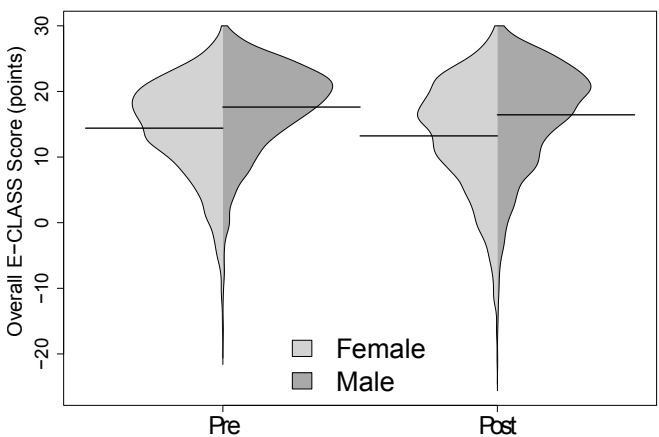}
\caption{Violin plots showing the distribution of overall pre- and postinstruction E-CLASS scores for men and women in the full, aggregate data set ($N=7013$).  Solid lines indicate the mean for each distribution.  Differences between the distributions for men and women are statistically significant for both pre- and post-tests (Mann-Whitney U, $p<<0.01$).  }\label{fig:overallAgg}
\end{figure}

\begin{figure*}
\includegraphics[width=\linewidth]{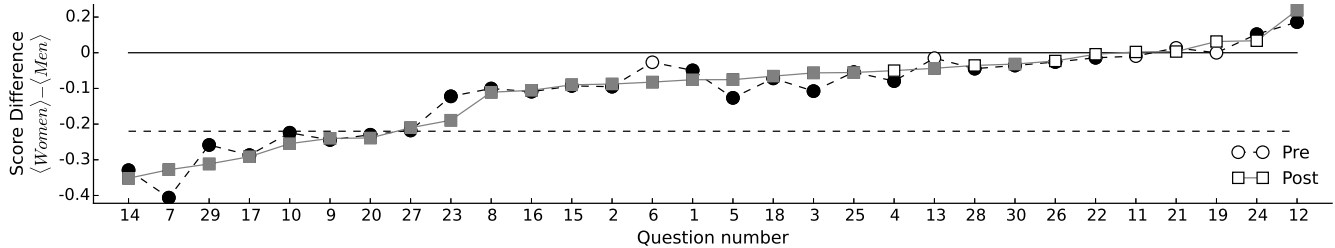}
\caption{Sorted plot of the difference between the average scores (points) of women and men on each item of the E-CLASS. Zero difference is marked by the solid horizontal line, and statistically significant differences are denoted by filled markers.  The dotted horizontal line marks the point below which the differences between men's and women's scores represent moderate effect sizes ($d>0.3$).  See Ref.\ \cite{ECLASSwebsite} for full list of question prompts. }\label{fig:byItem}
\end{figure*}

To determine if there are gender differences in students' performance on the E-CLASS, we first examine overall E-CLASS scores pre- and postinstruction for men and women.  As shown in Table \ref{tab:overallAgg}, there was a statistically significant gap between men and women's overall scores, and the magnitude of this gap represents a moderate effect size \cite{cohen1988d} with women scoring lower.  For the remainder of the paper, we will refer to gaps like this one as, for example, a statistically significant, moderate gap, where `moderate' here refers to the magnitude of the effect size.  The distributions of pre- and postinstruction scores for men and women are given in Fig.\ \ref{fig:overallAgg}.  

An examination of students' scores by item (Fig.\ \ref{fig:byItem}) shows that the gap between women and men's scores was small and relatively uniform across items.  The gender gap is statistically significant for 25 items preinstruction and 22 items postinstruction (Holm-bonferroni \cite{holm1979hb} corrected $p<0.05$).  With the exception of one item on the post-test with a statistically significant gap and two on the pretest, men outperformed women.  The magnitude of the gender gap was small ($d\le0.3$) for the majority of items ($N_{items}=23$) and moderate for the rest ($0.3<d<0.4$, $N_{items}=7$).  No obvious trend emerged in the content of these seven questions that might suggest why they resulted in larger gender differences.  

\begin{table}
\caption{Overall E-CLASS scores (points) for men and women in the full, aggregate data set ($N=7013$) on both the pre- and post-tests.  Significance indicates the statistical significance of the difference between women and men's scores.  }\label{tab:overallAgg}
\begin{ruledtabular}
   \begin{tabular}{ l r r r r}
       &Women& \hspace{4mm}Men& \hspace{2mm} Significance & \hspace{2mm} Effect Size\\
      \hline
      N & 2751 & 4262 & - & - \\
      Pre & 14.4 & 17.6 & $p\ll0.01$ & $d=-0.5$ \\
      Post \hspace{2mm} & 13.2 & 16.4 & $p\ll0.01$ & $d=-0.4$ \\
   \end{tabular}
\end{ruledtabular}
\end{table}

\begin{table}[b]
\caption{Formula for the four metrics for learning gain used here.  In some cases the formula has been generalized to account for the fact that the minimum E-CLASS score is $-30$ points rather than $0$ points.   }\label{tab:gains}
   \begin{tabular}{ p{2.3cm} c}
      \hline
      \hline
       Gain & Equation\\
      \hline
      Normalized change & $c=\begin{cases}\sfrac{(post-pre)}{(max-pre)} \text{  if } post>pre \\0 \hspace{28mm} \text{  if }  post=pre \\ \sfrac{(post-pre)}{(pre - min)}  \text{  if } post<pre\\ \end{cases}$ \\ \hline
      Hake gain & $\langle\langle g \rangle\rangle =(\langle post\rangle-\langle pre\rangle)/(max-\langle pre\rangle)$\\ \hline
      Avg. absolute gain & $g_{abs}=(post-pre)/(max-min)$\\ \hline
      Percent increase over pretest & $g_{\%}=(post-pre)/(pre-min)$\\
      \hline
      \hline
   \end{tabular}
\end{table}

\begin{figure}[b]
\includegraphics[width=\linewidth]{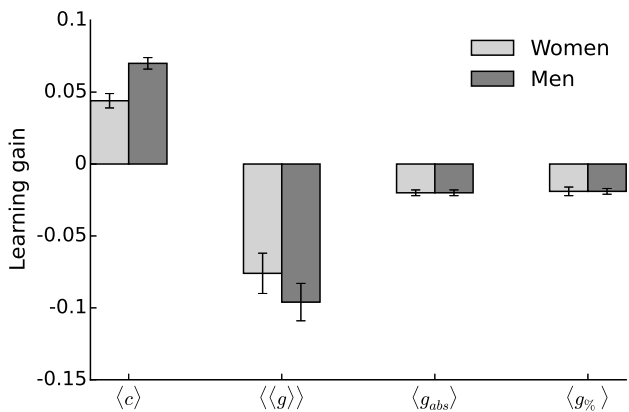}
\caption{Learning gains on each of the four different metrics for gain given in Table \ref{tab:gains}.  From left to right these are: average normalized change $\langle c\rangle$; Hake gain $\langle\langle g \rangle\rangle$; average absolute gain $\langle g_{abs}\rangle$; and average percent increase over pretest $\langle g_{\%}\rangle$.  Error bars represent the standard error of the mean.  The difference between the gains for men and women is statistically significant only in the case of normalized change $\langle c\rangle$ (independent sample t-test, $p\ll0.01$).  }\label{fig:gain}
\end{figure}

In addition to looking at raw pre- and postinstruction scores, it is also standard practice in the gender gap literature to examine some measure of gain as a proxy for how much students' understanding or attitudes changed over the course.  This change is often interpreted as the impact of instruction.  For our purposes, we might examine gain for two related reasons: to determine if instruction differentially benefits one gender more than the other, and to see if women make similar gains to those of men despite their lower preinstruction scores.   Consistent with the recommendations from Day \emph{et al.} \cite{day2016gender}, we calculate and compare gains from multiple common measures of learning gain including normalized change $\langle c\rangle$, Hake's normalized gain $\langle\langle g \rangle\rangle$, average absolute gain $\langle g_{abs}\rangle$, and percent increase over pretest $\langle g_{\%}\rangle$.  These four measures of gain are summarized in Table \ref{tab:gains}.

Fig.\ \ref{fig:gain} presents the results from each of these four metrics of gain.   In all cases, the magnitude of the gain was small, but statistically significant; however, both the magnitude and sign of the gain depended on the metric being used.  In particular, the average normalized change showed a positive gain despite the negative shift in raw score.  This is due to the skewed nature of the E-CLASS overall score distribution (Fig.\ \ref{fig:overallAgg}), which results in a suppression in the magnitude of negative gains relative to positive gains even for shifts of the same magnitude.  Average normalized change was also the only metric to result in a statistically significant difference between the gains of men and women (independent sample t-test, $p\ll0.01$).  

The inconsistency in the magnitude and sign of the gain, as well as the statistical significance of the difference in gain between men and women across different measures makes these results difficult to interpret.  This inconsistency between different measures of learning gain was also encountered by Day \emph{et al.} \cite{day2016gender} when characterizing the gender gap on another laboratory assessment.  In response to this issue, Day \emph{et al.} recommend shifting emphasis from examining learning gains to comparing postinstruction scores after controlling for preinstructions differences.  Analysis of covariance (ANCOVA) is one statistical method that allows us to control for multiple factors when looking at postinstruction means.  Sec.\ \ref{sec:stdemResults} identifies additional factors that should be controlled for in this comparison, and Sec.\ \ref{sec:ancovaResults} reports the results of an ANCOVA on data from E-CLASS.

\subsection{\label{sec:stdemResults}Gender differences in student subpopulations}

Up to this point, we have focused on identifying gender differences in the full, aggregate E-CLASS data set; however, there is significant variability in the types of courses represented in this data set, as well as the student populations of those courses (See Table \ref{tab:courseTypes} and Table \ref{tab:major}).  The gender gap may be similarly variable across different course types and student subpopulations.  For example, FY and BFY courses are often distinct in terms of class size, physics content, and complexity of equipment.  Table \ref{tab:courseLevelAvg} presents overall average scores for men and women for students in FY and BFY courses separately.  While the gender gap remained statistically significant both pre- and postinstruction in both the FY and BFY subpopulations, the size of the gap decreased from a moderate effect size in the FY to a small effect size in the BFY.  Additionally, both men and women in the BFY population scored significantly higher than those in the FY population ($p\ll0.01$).  

\begin{table}[b]
\caption{Overall E-CLASS scores (points) for men and women on both the pre- and post-tests for the FY and BFY student populations separately.  Significance indicates the statistical significance of the difference between women and men's scores.} \label{tab:courseLevelAvg}
\begin{ruledtabular}
   \begin{tabular}{ l l r r r r}
      Level & &Women& Men& Significance & Effect Size\\
      \hline
      &N & 2452 & 3040 & - & - \\
      FY & Pre & 14.1 & 17.3 & $p\ll0.01$ & $d=-0.5$ \\
      &Post & 12.8 & 15.7 & $p\ll0.01$ & $d=-0.4$ \\
      \hline
      &N & 299 & 1222 & - & - \\
      BFY &Pre & 17.3 & 18.4 & $p\ll0.01$ & $d=-0.2$ \\
      &Post & 17.1 & 18.2 & $p\ll0.01$ & $d=-0.2$ \\
   \end{tabular}
\end{ruledtabular}
\end{table}

Student major is another factor that may interact with the gender gap.  Physics majors, in particular, are a self-selected population that may exhibit different trends than the overall population.  The breakdown of students scores by major is given in Table \ref{tab:majorAvg}.  Here, we have focused specifically on the distinction between physics and non-physics majors, where non-physics includes all students not declared as physics or engineering physics majors.  Table \ref{tab:majorAvg} shows a statistically significant gap in the pre- and postinstruction scores for both physics and non-physics majors.  However, while the gap for non-physics majors was of moderate size, the gap for physics majors was of small effect size.  Additionally, both men and women who are physics majors scored significantly higher than students who are non-physics majors ($p\ll0.01$).  

\begin{table}
\caption{Overall E-CLASS scores (points) for men and women on both the pre- and post-tests for the physics and non-physics student populations separately.  Significance indicates the statistical significance of the difference between women and men's scores.} \label{tab:majorAvg}
\begin{ruledtabular}
   \begin{tabular}{ l l r r r r}
      Major & &Women& \hspace{2mm}Men& \hspace{1mm} Significance & \hspace{1mm} Effect Size\\
      \hline
      &N & 2476 & 3062 & - & - \\
      Non-physics &Pre & 14.0 & 17.0 & $p\ll0.01$ & $d=-0.4$ \\
      &Post  & 12.6 & 15.4 & $p\ll0.01$ & $d=-0.3$ \\
      \hline
      &N & 275 & 1200 & - & - \\
      Physics &Pre & 18.2 & 19.1 & $p<0.01$ & $d=-0.2$ \\
      &Post & 18.6 & 19.2 & $p=0.04$ & $d=-0.09$\\
   \end{tabular}
\end{ruledtabular}
\end{table}

These results suggest that, as predicted, there was significant variation in the size of the gender gap for some subpopulations of students.  However, course level and distribution of student major are not independent factors.  For example, BFY courses are far more likely to have a majority of physics majors.  To more clearly characterize the variations in students scores with respect to course level and major, we must examine these factors intersectionally.  Table \ref{tab:BFYmajor} provides the breakdown of students pre- and postinstruction scores by major for the BFY students only.  Similar to the findings for majors in the aggregate data, BFY physics majors still scored significantly higher than BFY non-physics majors ($p\ll0.01$).  However, the gender gap in both pre- and postinstruction scores was statistically significant for BFY nonphysics majors only; there was no significant difference between the scores of men and women for BFY physics majors.  

\begin{table}[b]
\caption{Overall E-CLASS scores (points) for men and women on both the pre- and post-tests for the BFY physics and BFY non-physics student populations separately.  Significance indicates the statistical significance of the difference between women and men's scores.} \label{tab:BFYmajor}
\begin{ruledtabular}
   \begin{tabular}{ p{1.7cm} l r r r r}
      Major & &Women & \hspace{2mm}Men& \hspace{1mm} Significance & \hspace{1mm} Effect Size\\
      \hline
      \multirow{3}{1.7cm}{BFY, non-physics}&N & 102 & 338 & - & - \\
      &Pre & 14.8 & 17.4 & $p\ll0.01$ & $d=-0.4$ \\
      &Post  & 13.0 & 15.5 & $p<0.01$ & $d=-0.3$ \\
      \hline
      \multirow{3}{1.7cm}{BFY, physics}&N & 197 & 884 & - & - \\
      &Pre & 18.6 & 18.8 & $p=0.4$ &  \\
      &Post & 19.1 & 19.2 & $p=0.5$ &  \\
   \end{tabular}
\end{ruledtabular}
\end{table}

The disappearance of the gender gap in the BFY physics major data was not replicated in the population of FY students, where a statistically significant, moderate gender gap persisted even when the data were disaggregated by major.  This finding suggests that there may be interactions between gender, major, and course level in these data.  Moreover, the preinstruction gender gap in E-CLASS scores makes it difficult to clearly interpret differences in postinstruction scores.  To clearly determine the size and significance of the gender gap for different subpopulations, we need to account for the potential impact of multiple factors simultaneously.  The next section addresses this issue using an analysis of covariance.

\subsection{\label{sec:ancovaResults}Analysis of covariance}

The previous sections identified several factors that correlated with students' postinstruction scores on the E-CLASS, including students' preinstruction scores, major, course level, and gender.  These factors, however, do not necessarily represent independent variables.  For example, Sec.\ \ref{sec:stdemResults} showed that the impact of one factor (e.g., gender) on postinstruction scores may depend on another factor (e.g., course level).  To disentangle the relationships between these different variables and explore the relationship between gender and postinstruction scores, we performed an ANCOVA (analysis of covariance) \cite{wildt1978ancova}.  ANCOVA is a statistical method for comparing the difference between population means while accounting for the variance associated with other factors.  In this case, we want to determine if the difference between postinstruction means for men and women is statistically significant after accounting for the impact of preinstruction scores, major, and course level.  Only students for whom we had data for both major and gender, in addition to matched E-CLASS data, were included in the ANCOVA analysis ($N=6968$).

We initially performed a 4-way ANCOVA that included preinstruction scores as a covariate in addition to the three categorical variables: major, course level, and gender.  However, in order to reliably interpret the impact of each of these variables individually, we must first determine if there were any statistically significant interactions between them.  The presence of such an interaction would violate one of the assumptions of an ANCOVA (i.e., homogeneity of the regression slopes \cite{wildt1978ancova, day2016gender}).  To test this, we included in the ANCOVA all possible interaction terms, and consistent with the results in Sec.\ \ref{sec:stdemResults}, the 4-way ANCOVA revealed a significant interaction between level and major (F-test \cite{wildt1978ancova}, $p=0.04$).  The existence of this interaction means that the variables course level and major must be analyzed independently.    A summary of the main findings of the separate ANCOVAs described in the remainder of this section is given in Table \ref{tab:ancova}.  

To analyze the significance of gender as predictors of postinstruction scores for course level and major separately, we first split the data by course level and ran separate 3-way ANCOVAs for each level.  The 3-way ANCOVA included preinstruction scores, major, and gender as variables.  We found that among FY students the adjusted postinstruction mean for men was significantly higher than the adjusted mean for women (F-test, $p<0.01$); however, among BFY students, the adjusted means for men and women were the same (F-test, $p=0.6$).  Preinstruction score and major were statistically significant predictors for both FY and BFY populations ($p\ll0.01$). Thus, after adjusting for the variance associated with preinstruction score and major, gender was a significant predictor of students' postinstruction E-CLASS score only for students in the FY courses (see Table \ref{tab:ancova}).    

\begin{table}
\caption{Impact of each categorical variable on postinstruction means as adjusted by the 3-way ANCOVAs.  Adjusted means for each variable are calculated controlling for preinstruction score and the other relevant categorical variable, as described in the text.  A difference between group means is indicated only when that difference was statistically significant.  Here, $\langle P\rangle$ is the predicted postinstruction mean for physics students, and similarly for non-physics students $\langle NP\rangle$, men $\langle M\rangle$, women $\langle W \rangle$, BFY students $\langle BFY\rangle$, and FY students $\langle FY \rangle$.   } \label{tab:ancova}
\begin{ruledtabular}
   \begin{tabular}{ l c c c}
      & \multicolumn{3}{c}{Catagorical Variable} \\
      Group &  Course level \hspace{4mm} & Gender &Major\\
      \hline
      Physics 	     & $\langle BFY\rangle>\langle FY\rangle$ & $\langle M\rangle=\langle W\rangle$ & - \\
      Non-physics & $\langle BFY\rangle=\langle FY\rangle$ & $\langle M\rangle>\langle W\rangle$ & - \\
      \hline
      FY  & - & $\langle M\rangle>\langle W\rangle$ & $\langle P\rangle>\langle NP\rangle$\\
      BFY & - & $\langle M\rangle=\langle W\rangle$ & $\langle P\rangle>\langle NP\rangle$ \\
   \end{tabular}
\end{ruledtabular}
\end{table}

The significance of course level as a predictor of postinstruction scores was determined by splitting the data by major and running separate 3-way ANCOVAs for each major.  This time the 3-way ANCOVA included preinstruction scores, gender, and course level as variables.  For non-physics majors, the adjusted postinstruction mean for men was significantly higher than the adjusted mean for women (F-test, $p<0.01$); however, the same trend did not hold for physics majors (F-test, $p=0.9$).  Alternatively with respect to course level, the adjusted means for BFY physics majors was signficantly higher than the adjusted mean for FY physics majors (F-test, $p<0.01$); but for non-physics majors, the adjusted means for FY and BFY were the same (F-test, $p=0.9$).  Thus, after adjusting for the variance associated with preinstruction E-CLASS score, gender was a significant predictor of postinstruction performance only for non-physics majors, and course level was a significant predictor only for physics majors (see Table \ref{tab:ancova}).

\section{\label{sec:discussion}Summary and Discussion}

We analyzed a large, national data set of student responses to a laboratory-focused assessment --- the E-CLASS --- to identify any significant gender differences in these responses.  Informed by the broader literature around performance gaps in physics, we not only examined students' performance with respect to gender, but also with respect to other student and course demographics (e.g., major and course level) that may have contributed to the variance in overall E-CLASS score.  By examining the raw pre- and postinstruction E-CLASS means for students at the intersections of gender, course level, and major, we found that the size of the gender gap varied significantly, and in some cases even disappeared, for specific subpopulations (e.g., BFY physics majors).  This finding was also supported by the results of an ANCOVA (summarized in Table \ref{tab:ancova}), which examined the difference between postinstruction means on the E-CLASS while accounting for the variance associated with preinstruction scores, course level, major, and gender simultaneously.  The ANCOVA showed that when looking at different course levels separately, gender was a statistically significant predictor of postinstruction performance only in the first-year courses.  Additionally, when looking at majors separately, gender was a significant predictor only for non-physics majors.  For researchers interested in investigating gender or performance gaps, our findings underscore the importance of considering sources of within-group variance when comparing performance between groups of students.

Together, these results suggest that some factor (or set of factors) resulted in differentially lower than expected E-CLASS scores for FY women who are non-physics majors relative to men who are non-physics majors.  This factor (or factors) did not result in a similar suppression of scores for FY women who are physics majors.  This, combined with previous research linking students' attitudes, confidence, and epistemologies with their interest and persistence in the major, suggest that the population of FY non-physics women may be a key population for instructors and researchers to consider when working to improve students' attitudes about physics, as well as the persistence and recruitment of women into the physics major.  However, an important limitation of this work is that the nature of the factor(s) that caused the reduction in the scores of FY, nonphysics women cannot be determined from these analyses.  Moreover, we cannot determine why this effect does not persist into the BFY population of women.  One potential hypotheses might be that this effect was caused by a differentially positive (or less negative) impact of FY or BFY instruction on women relative to men.  Alternatively, the disappearance of the gender gap in the BFY courses could be a result of a differential selection effect as only a subset of women persist through the physics curriculum.  It is also possible that this finding is driven by an entirely different source or a combination of these and other sources.  

In addition to the lack of data that can speak to a causal mechanism for the appearance and disappearance of the gender gap in E-CLASS data, there are several additional limitations of this work.  Our data set is extensive and spans a large number of institutions, courses, and student populations; however, it is neither comprehensive nor randomly selected.  For example, there are only a few 2-year colleges in our data.  Moreover, the instructors for the courses in our data set generally chose to use E-CLASS without external pressure, and thus represent a self-selected group.  Additionally, we focused here on a specific subset of potential variables that might impact the gender gap in postinstruction E-CLASS scores (i.e., major, course level, and preinstruction scores).  These variables were selected based on preliminary analysis of the data and our own experience, which suggested they could account for significant amounts of the variance in postinstruction E-CLASS scores.  However, there are other factors that might also correlate with gender differences in students' epistemologies, affect, and confidence with respect to experimental physics including, for example, high school laboratory experiences, course structure, pedagogy, or participation in undergraduate research experiences.  Indeed, some of these factors may have contributed to the persistent gender gap observed in students preinstruction E-CLASS scores.  

While awareness of the existence of and variations in gender differences in E-CLASS scores is important for instructors, the current work does not provide insight into instructional strategies that might address the gap.  Ongoing work with this data set looks for variations in gender differences based on instructor's use of different pedagogical techniques and types of classroom activities.  Future work around gender differences on the E-CLASS could include qualitative investigations targeted at understanding the causal mechanism behind the persistence of the gender gap in first-year courses.  Additionally, longitudinal studies following cohorts of students through multiple laboratory courses could be used to determine whether there is a differential selection effect between men and women that accounts for the disappearance of the postinstruction gender gap in beyond-first-year lab courses.  While longitudinal data is notoriously difficult to collect, we continue to aggregate data from CU that may shed light on this question in the future.  Moreover, while our findings indicated that the gender gap in postinstruction scores was often partially or completely explained by factors other than gender, the gap in preinstruction E-CLASS scores persists across almost all subpopulations.  Additional quantitative and qualitative analysis of students' incoming experience and epistemology will be necessary to understand this preinstruction gap and determine its significance for the recruitment and persistence of women in the physics major.

\begin{acknowledgments}
This work was funded by the NSF-IUSE Grant DUE-1432204 and NSF Grant PHY-1125844.  Particular thanks to Dimitri Dounas-Frazer and to the members of PER@C for all their help and feedback.  
\end{acknowledgments}

\bibliography{master-refs-ECLASS-2_16}

\end{document}